\documentstyle[12pt,epsfig]{article} 
\title{ Citations and the Zipf-Mandelbrot's law}
\author { Z.~K.~Silagadze 
\vspace*{3mm} \\ Budker Institute of Nuclear Physics,  630 090,
Novosibirsk, Russia }
\date{}

\begin{document}
\large
\maketitle

\begin{abstract}
A curious observation was made that the rank statistics of scientific
citation numbers follows Zipf-Mandelbrot's law. The same pow-like behavior
is exhibited by some simple random citation models. The observed 
regularity indicates not so much the peculiar character of the
underlying (complex) process, but more likely, than it is usually assumed,
its more stochastic nature.
\end{abstract}

\newpage
\section{Introduction}
Let us begin with an explanation as to what is Zipf's law. If we assign ranks
to all words of some natural language according to their frequencies in some 
long text (for example the Bible), then the resulting frequency-rank
distribution follows a very simple empirical law
\begin{equation}
f(r)=\frac{a}{r^\gamma}
\label{eq1} \end{equation}
\noindent with $a\approx 0.1$ and $\gamma\approx 1$. This was observed by 
G.~K~Zipf for many languages long time ago \cite{1,2}. More modern studies
\cite{3} also confirm a very good accuracy of this rather strange regularity.

In his attempt to derive the Zipf's law from the information theory,
Mandelbrot \cite{4,5} produced a slightly generalized version of it:
\begin{equation}
f(r)=\frac{p_1}{(p_2+r)^{p_3}},
\label{eq2} \end{equation}
\noindent $p_1,p_2,p_3$ all being constants.

The same inverse pow-law statistical distributions were found in 
embarrassingly different situations (For reviews see \cite{6,7}). 
In economics,
it was discovered by Pareto \cite{8} long ago before Zipf and states that 
incomes of individuals or firms are inversely proportional to their rank. 
In less formal words \cite{9}, ``most success seem to migrate to those people
or companies who already are very popular''. In demography \cite{2,10,11},
city sizes (populations) also are pow-like functions of cities ranks. The same
regularity reveals itself in the distributions of areas covered by satellite 
cities and villages around huge urban centers \cite{12}.

Remarkably enough, as is claimed in \cite{13}, in countries such as former 
USSR and China, where natural demographic process were significantly 
distorted, city sizes do not follow Zipf's law!

Other examples of zipfian behavior is encountered in chaotic dynamical 
systems with multiple attractors \cite{14}, in biology \cite{15}, ecology
\cite{16}, social sciences and etc. \cite{17}.

Even the distribution of fundamental physical constants, according to 
\cite{18}, follows the inverse power law!

The most recent examples of Zipf-like distributions are related to the
World Wide Web surfing process \cite{19,20}.

You say that all this sounds like a joke and looks improbable? So did I
when became aware of this weird law from M.~Gell-Mann's book ``The Quark
and the Jaguar'' \cite{21} some days ago. But here are the distribution
of first 50 USA largest cities according to their rank \cite{22}, fitted by
Eq.2:
\begin{figure}[htb]
  \begin{center}
\mbox{\epsfig{figure=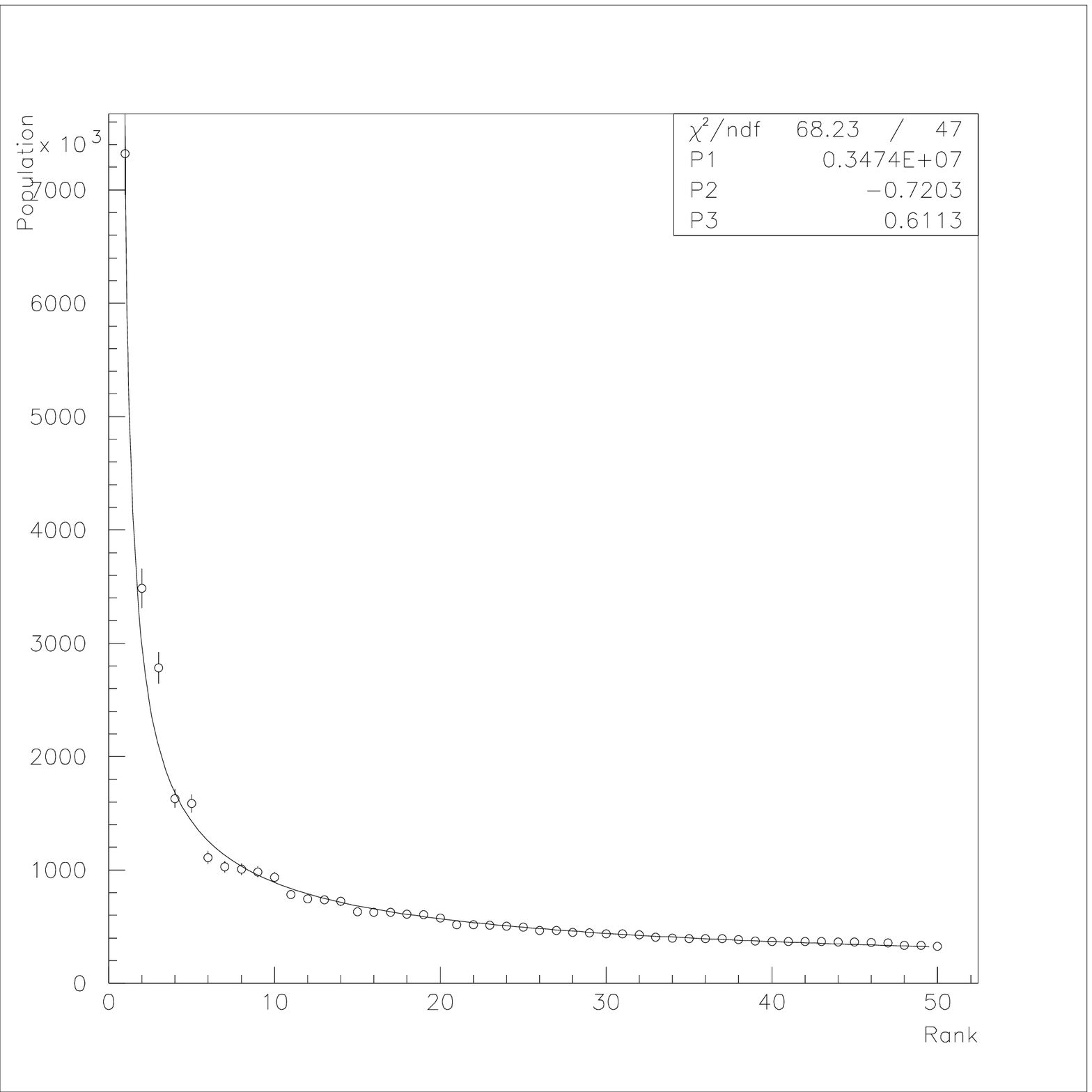
                               ,  height=14.0cm}}
   \end{center}
%\caption {USA largest cities}
\label{Fig1}
\end{figure}

\noindent The actual values of fitted parameters depend on the details 
of the fit. I assume (rather arbitrarily) 5\% errors in data.

Maybe it is worthwhile to remember here, the old story about a young priest 
who complains his father about having a very difficult theme for his first
public sermon -- virgin birth.

-- ``Look father'', he says, ``if some young girl from this town, becomes 
pregnant,
comes to you and says that this is because of Holy Spirit. 
Do you believe it?''

The father stays silent for a while, then answers:

--''Yes, son, I do. If the baby would be born, if he would be raised and if he
would live like the Christ''.

So, clearly, you need more empirical evidence to accept improbable things.
Here is one more, the list of the most populated countries \cite{23} fitted
by the Mandelbrot formula (\ref{eq2}):
\begin{figure}[htb]
  \begin{center}
\mbox{\epsfig{figure=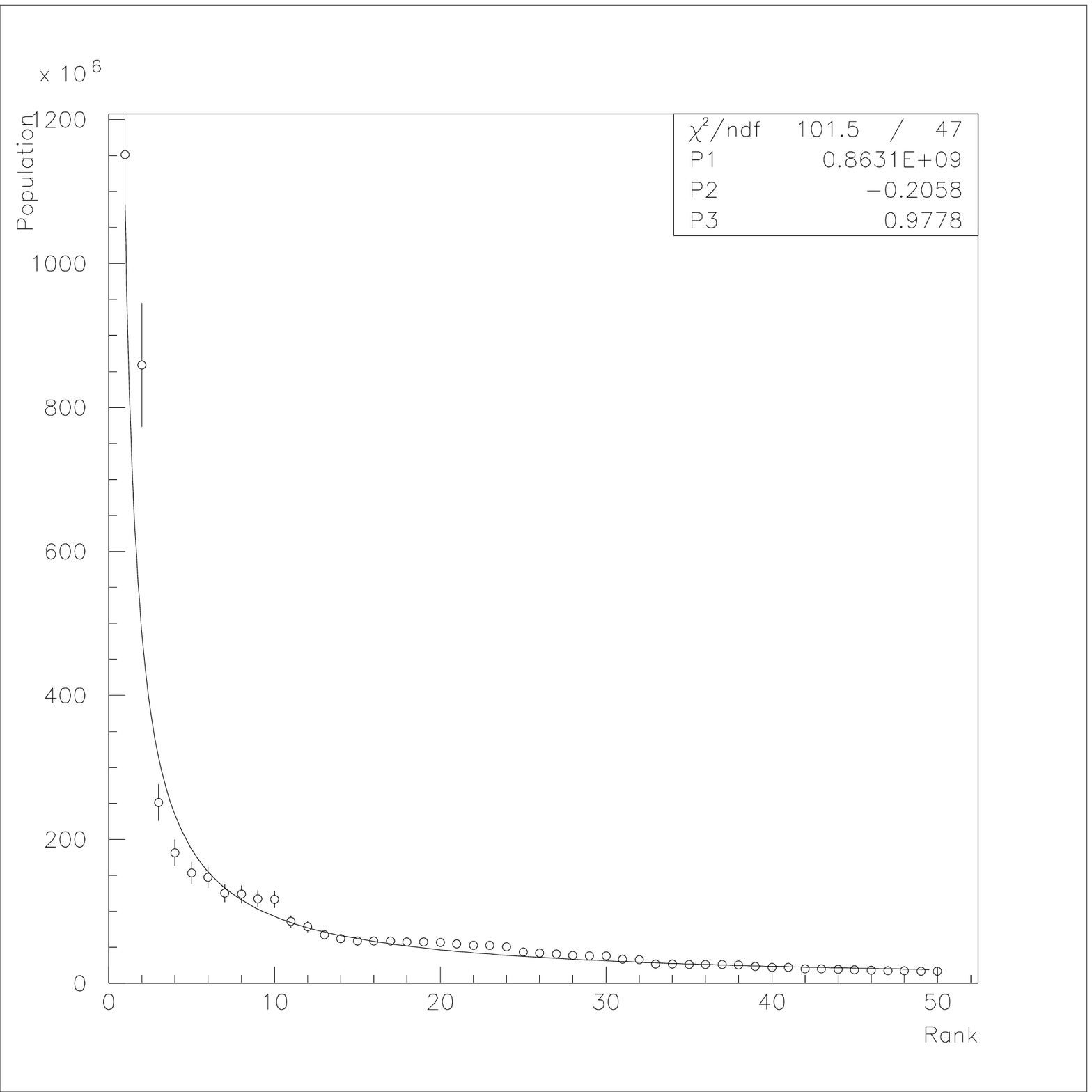
                               ,  height=14.0cm}}
   \end{center}
%\caption {Most populated countries}
\label{Fig2}
\end{figure}

\noindent Even more simple Zipfian $a/r$ parameterization will work in this
case fairly well!

\section{Fun with citations}
But all this was known long ago. Of course it is exciting to check its 
correctness personally. But more exciting is to find whether this rule still
holds in a new area. SPIRES database provides excellent possibility to check
scientific citations against Zipf-Malderbrot's regularity.

As I have been involved in this matters because of M.~Gell-Mann's book, 
my first try naturally was his citations itself. The results were 
encouraging: 

\begin{figure}[htb]
  \begin{center}
\mbox{\epsfig{figure=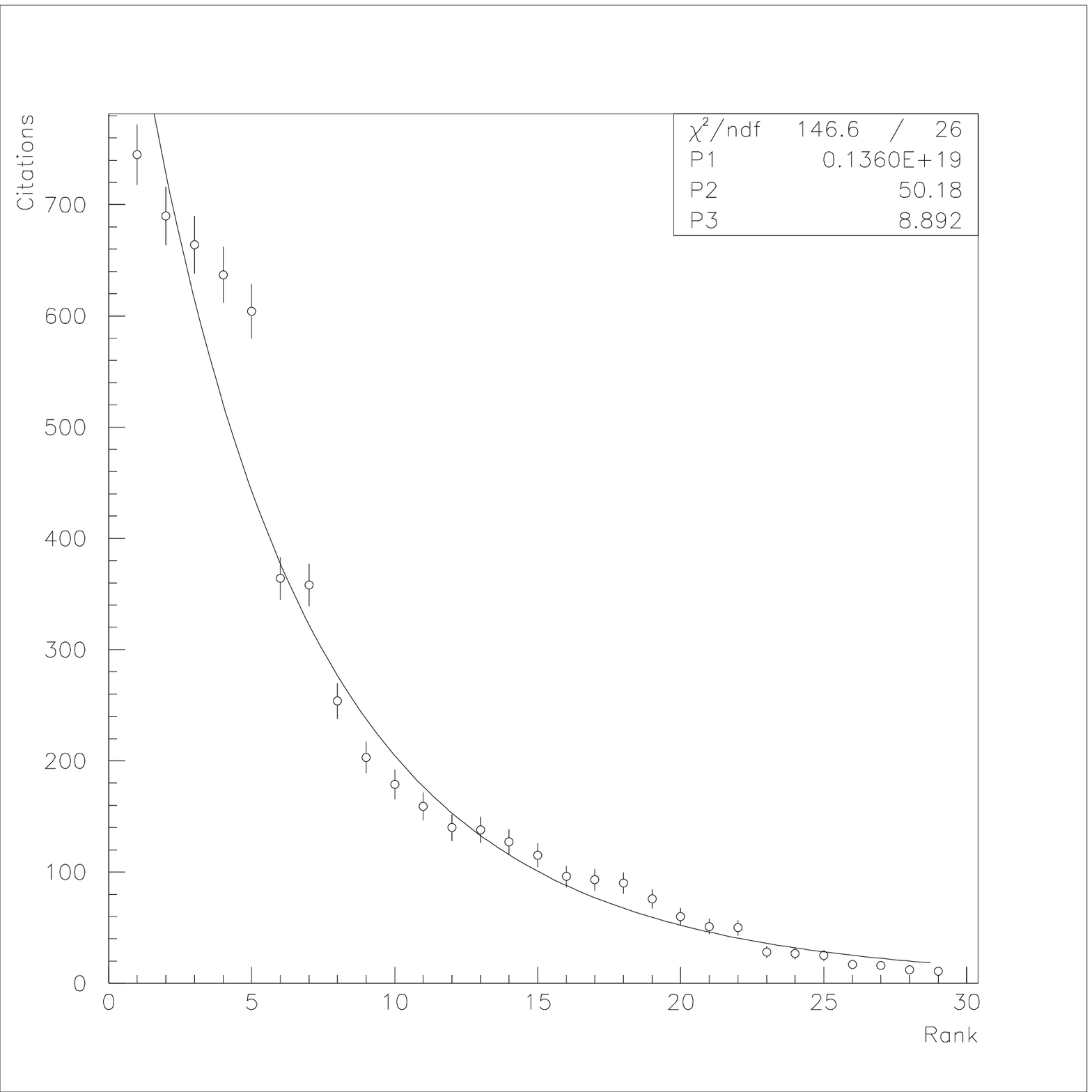
                               ,  height=14.0cm}}
   \end{center}
%\caption {Gell-Mann's citations}
\label{Fig3}
\end{figure}

But maybe M.~Gell-Mann is not the best choice for this goal. SPIRES is 
a rather novel phenomenon, and M.~Gell-Mann's many important papers were
written long before its creation. So they are purely represented in the
database. Therefore, let us try present day citation favorite E.~Witten.
Here are his 160 most cited papers according to SPIRES \cite{24} (Note once 
more that the values of fitted parameters may depend significantly on the 
details of the fit. In this and previous case I choose $\sqrt{N}$ as 
an estimate for data errors, not to ascribe too much importance to data 
points with small numbers of citations. In other occasions I assume 5\% 
errors. Needless to say, both choices are arbitrary): 
\newpage
\begin{figure}[htb]
  \begin{center}
\mbox{\epsfig{figure=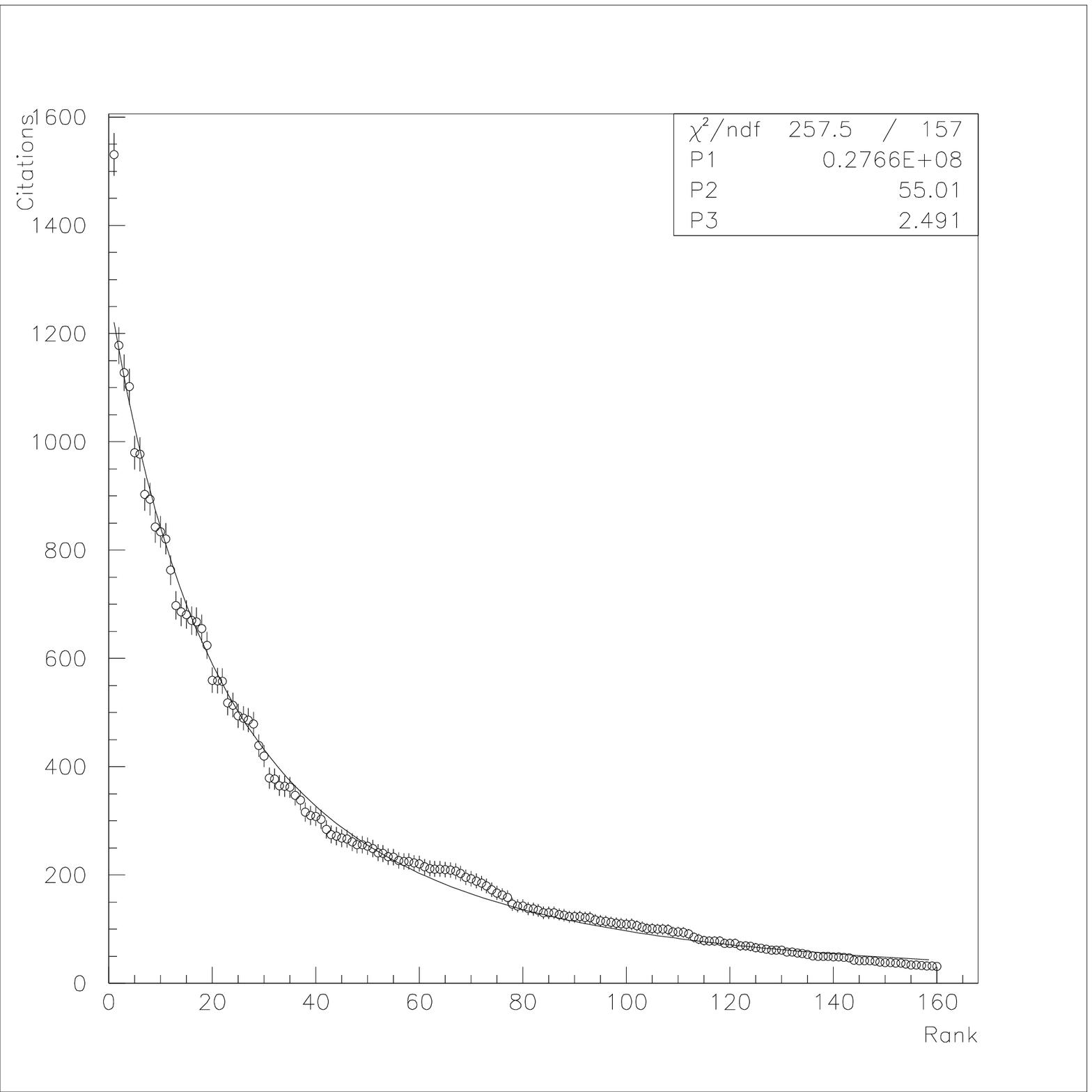
                               ,  height=15.0cm}}
   \end{center}
%\caption {Witten's citations}
\label{Fig4}
\end{figure}

You have probably noticed very big values of the prefactor $p_1$. 
Of course this is
related to the rather big values of other two parameters. We can understand 
big value of $p_2$ parameter as follows. The data set of individual 
physicist's papers are subset of more full data about all physicists. 
So we can
think of $p_2$ as being an average number of papers from other scientists
between two given papers of the physicists under consideration. Whether right
or not, this explanation gains some empirical support if we consider top
cited papers in SPIRES \cite{25} (Review of particle physics is excluded):
\newpage
\begin{figure}[htb]
  \begin{center}
\mbox{\epsfig{figure=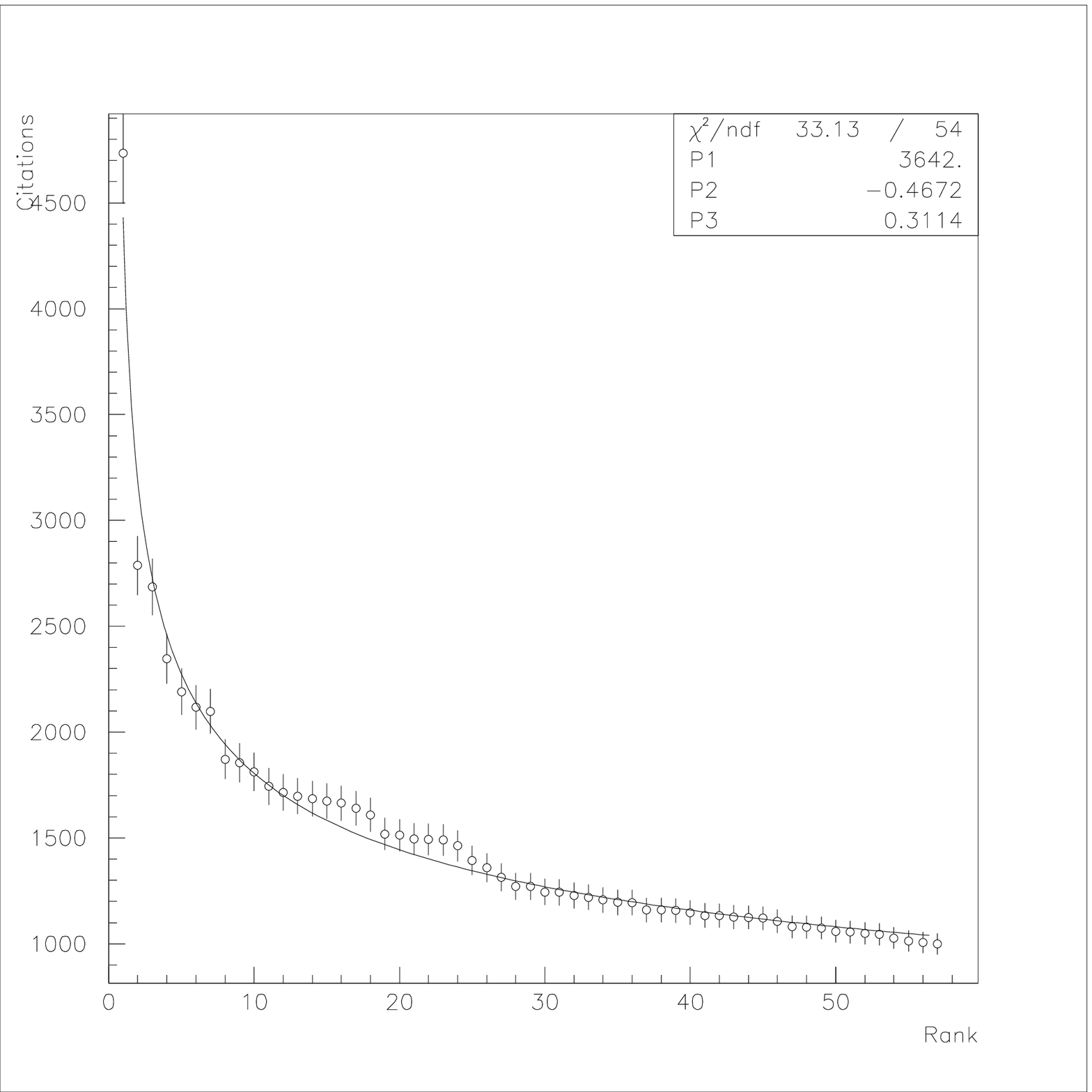
                               ,  height=15.0cm}}
   \end{center}
%\caption {Top cited papers}
\label{Fig5}
\end{figure}

\noindent As we see $p_2$ is fairly small now.

At last, it is possible to find the list of 1120 most cited physicists
(not only from the High Energy Physics) on the World Wide Web \cite{26}.
Again the Mandelbrot formula (\ref{eq2}) with $p_1=3.81\cdot 10^4,\;
p_2=10.7$ and $p_3=0.395$ gives an excellent fit. Now there are too
many points, making it difficult to note visually the differences between the
curve and data. In the figure that follows, we show this relative difference 
explicitly.
\newpage
\begin{figure}[htb]
  \begin{center}
\mbox{\epsfig{figure=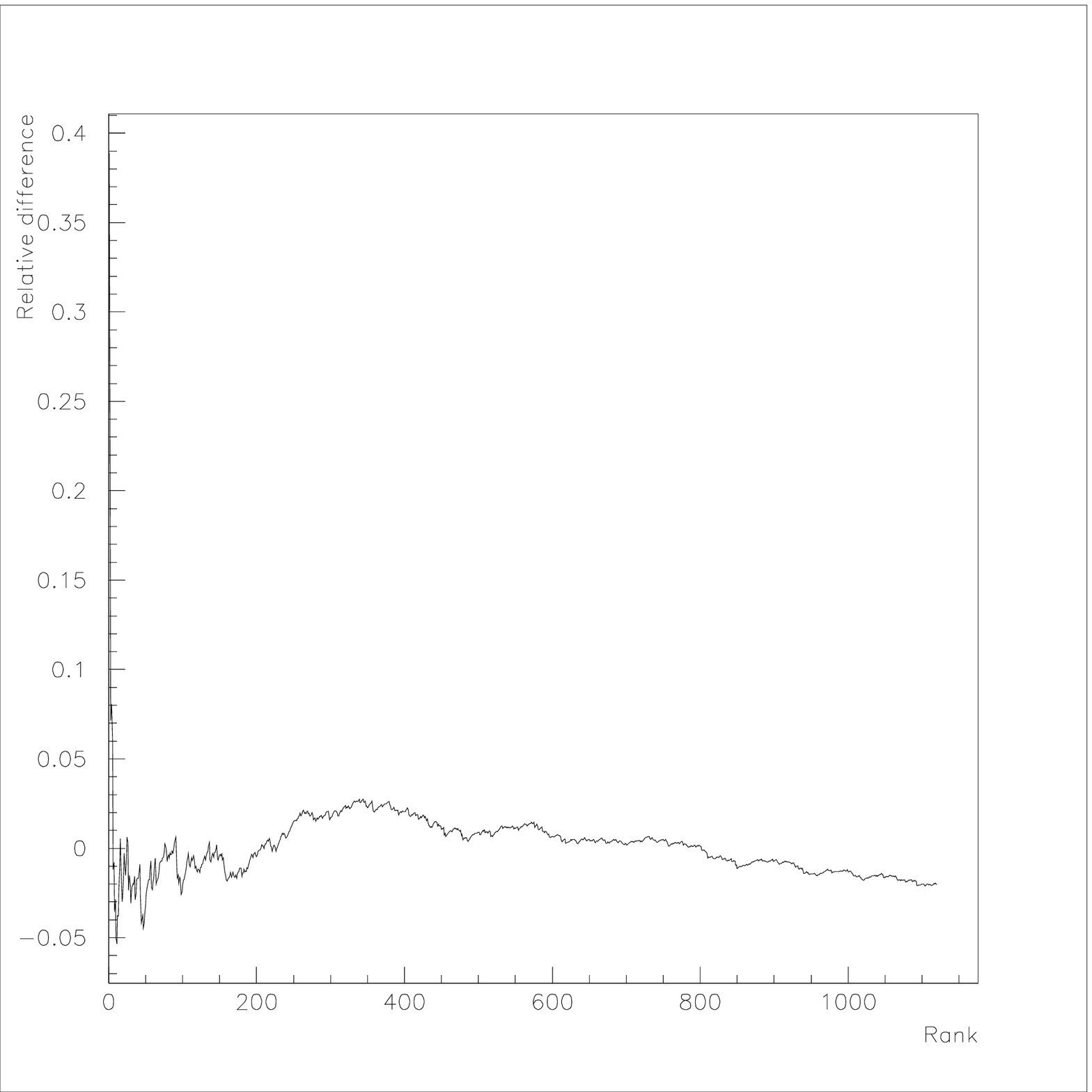
                               ,  height=12.0cm}}
   \end{center}
%\caption {1120 most cited papers}
\label{Fig6}
\end{figure}

For the most bulk of data the Mandelbrot's curve gives the precision better
than 5\%! 

You wonder why now $p_2$ is relatively high? I really do not know. Maybe
the list is still incomplete for his lower rank part. In any case, if you
take just the first 100 entries from this list, the fit results in
$p_1=2.1 \cdot 10^4,\; p_2=-0.09,\; p_3=0.271$. This example also shows
that actually the Mandelbrot's curve with constant $p_1,\; p_2,\; p_3$
is not as good approximation as one might judge from the above given 
histograms, because different parts of data prefer different values of
the Mandelbrot's parameters. 

\section{Any explanation?}
The general character of the Zipf-Mandelbrot's law is hypnotizing. We already
mentioned several wildly different areas where it was encountered. Can it be
considered as some universal law for complex systems? And if so, what is
the underlying principle which unifies all of these seemingly different
systems? What kind of principle can be common for natural languages, 
individual wealth distribution in some society, urban development, scientific
citations, and female first name frequencies distribution? The latter is
reproduced below \cite{27}:
\begin{figure}[htb]
  \begin{center}
\mbox{\epsfig{figure=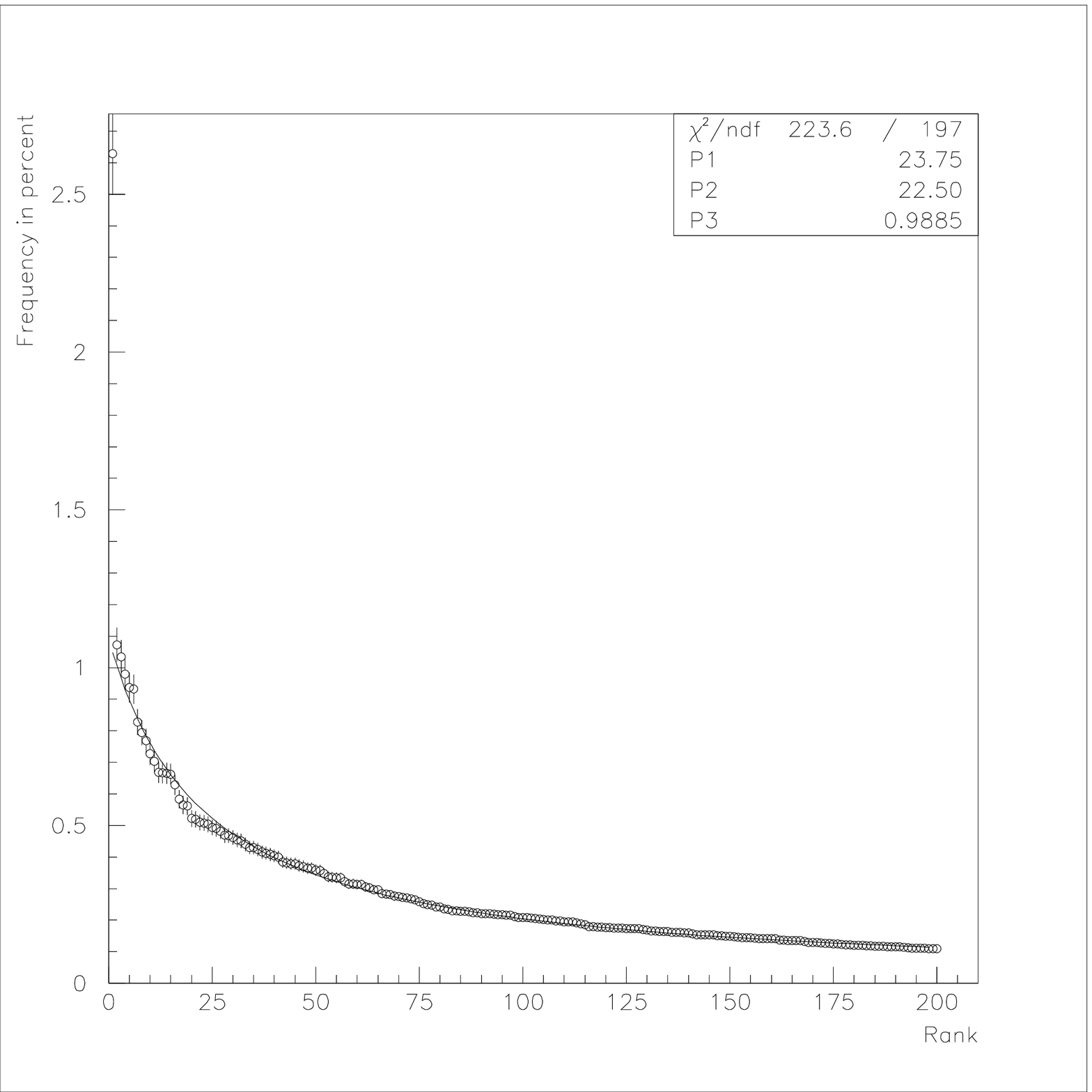
                               ,  height=14.0cm}}
   \end{center}
%\caption {Female first names}
\label{Fig7}
\end{figure}

Another question is whether the Mandelbrot's parameters $p_2$ and $p_3$ can
tell us something about the (complex) process which triggered the 
corresponding
Zipf-Mandelbrot distribution. For this goal an important issue is how to 
perform the fit (least square, $\chi^2$, method of moments \cite{20} or 
something else?). I do not have any answer to this question now. However
let us
compare the parameters for the female first name distribution from the above
given histogram and for the male first name distribution (data are taken from
the same source \cite{27}). In both cases $\chi^2$ fit was applied with 5\%
errors assumed for each point.
\begin{figure}[htb]
  \begin{center}
\mbox{\epsfig{figure=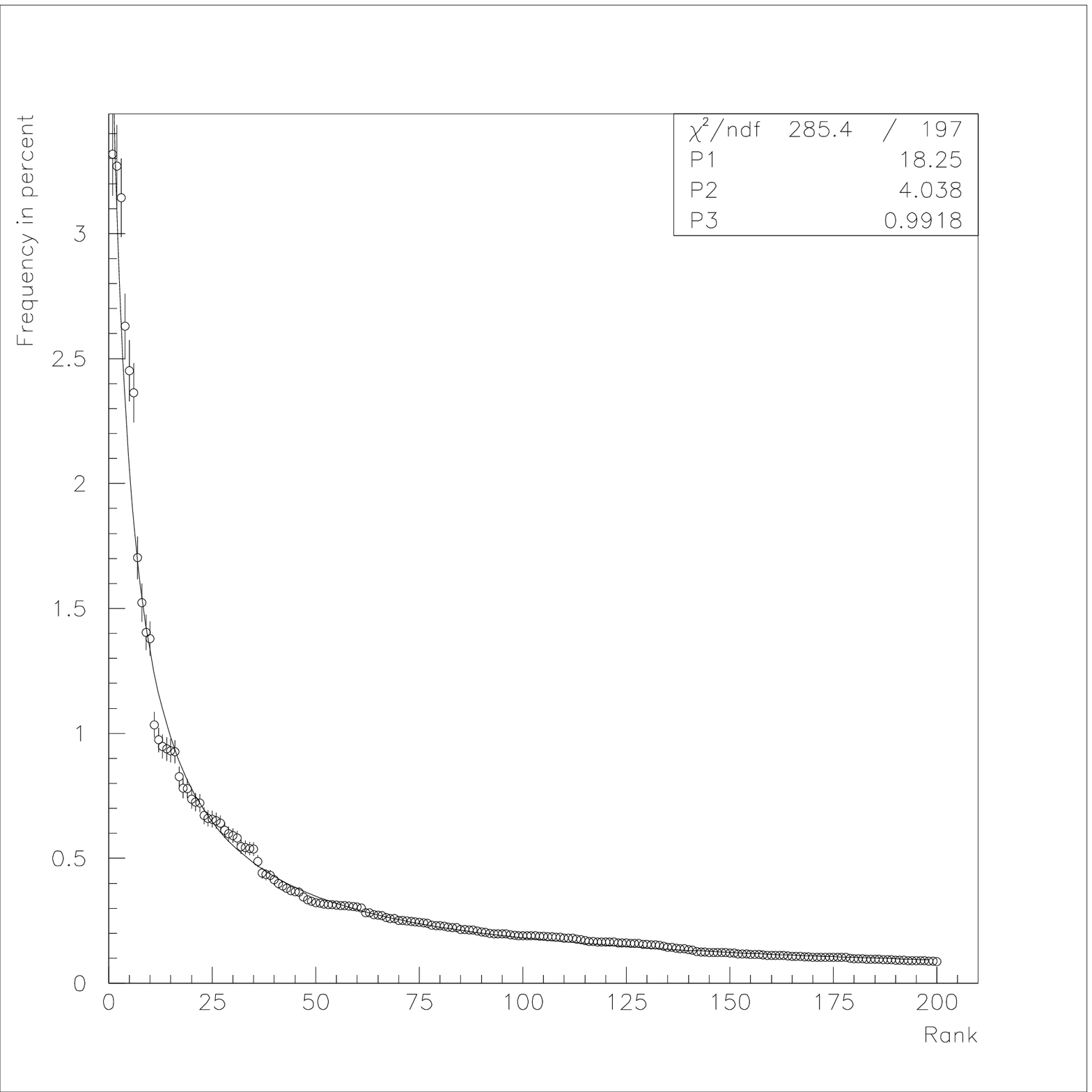
                               ,  height=14.0cm}}
   \end{center}
%\caption {Male first names}
\label{Fig8}
\end{figure}

The power-counting parameter $p_3$ is the same for both distributions, 
although the $p_2$ parameter has different values.

If you are fascinated by a possibility that very different complex systems
can be described by a single simple law, you maybe will be disappointed
(as was I) to learn that some simple stochastic processes can lead to very
same Zipfian behavior. Say, what profit will you have from knowing that some
text exhibits Zipf's regularity, if this gives you no idea the text was 
written by Shakespeare or by monkey? Alas, it was shown \cite{4,28,29,30} 
that  random texts (``monkey languages'') exhibit Zipf's-law-like word 
frequency distribution. So Zipf's law seems to be at least \cite{5}
``linguistically very shallow'' and \cite{29} ``is not a deep law in natural
language as one might first have thought''.

Two different approaches to the explanation of Zipf's law is very well
summarized in G.~Millers introduction to the 1965 edition of Zipf's book
\cite{1}: ``Faced with this massive statistical regularity, you have two
alternatives. Either you can assume that it reflects some universal property
of human mind, or you can assume that it reflects some necessary consequence
of the laws of probabilities. Zipf chose the synthetic hypothesis and 
searched for a principle of least effort that would explain the apparent
equilibrium between uniformity and diversity in our use of words. Most others
who were subsequently attracted to the problems chose the analytic hypothesis
and searched for a probabilistic explanation. Now, thirty years later, it 
seems clear that the others were right. Zipf's curves are merely one way 
to express a necessary consequence of regarding a message source as 
a stochastic process''.

Were ``others'' indeed right? Even in the realm of linguistics the debate is
still not over after another thirty years have passed \cite{31}. In the case
of random texts, the origin of the Zipf's law is well understood 
\cite{32,33}.
In fact such texts exhibit no Zipfian distribution at all, but log-normal
distribution, the latter giving in some cases a very good approximation to
the Zipf's law. So there is no doubt that simple stochastic (Bernoulli or
Markov) processes can lead to a Zipfian behavior. No dynamically nontrivial
properties (interactions and interdependence) is required at all from the 
underlying system. But it was also stressed in the literature \cite{34,13}
that this fact does not preclude more complex and realistic systems to
exhibit Zipfian behavior because of underlying nontrivial dynamics. In this
case, we can hope that the Zipf-Mandelbrot parameters will be meaningful and
can tell something about the system properties. Let us note that the 
rank-frequency distribution for complex systems is not always Zipfian. For
example, if we consider the frequency of occurrence of letters, instead of
words, in a long text, the empirical universal behavior, valid over 100
natural languages with alphabet sizes ranged between 14 and 60, is 
logarithmic \cite{35}
$$ f(r)=A-B\ln{r} \, $$
\noindent where $A$ and $B$ are constants. This fact, of course, is 
interesting by itself. It is argued in \cite{35} that both regularities
(zipfian and logarithmic) can have the common stochastic origin.

An interesting example of Zipf-Mandelbrot's parameters being useful and
effective, is provided by ecology \cite{36,37}. The exponent $p_3$ is related
to the evenness of the ecological community. It has higher values for 
``simple'' and lower values for ``complex'' systems. The parameter $p_2$ is
related to the ``diversity of the environment'' \cite{37} and serves as 
a measure of the complexity of initial preconditions.

The another pole in explanation of Zipf's law seeks some universal principle
behind it, such as ``least effort'' \cite{2}, ``minimum cost'' \cite{4},
``minimum energy'' \cite{38} or ``equilibrium'' \cite{39}. The most 
impressive and, as the above ecological example shows, fruitful explanation
is given by B.~Mandelbrot \cite{5,40} and is based on fractals and 
self-similarity.

As we see, the suggested explanations are almost as numerous as the observed
manifestations of this universal pow-like behavior. This probably indicates
that some important ingredient in this regularity still escapes 
to be grasped.
As M.~Gell-Mann concludes \cite{21} ``Zipf's law remains essentially
unexplained''.

\section{The almighty chance}
If monkeys can write texts they can make citations too! So let us imagine
the following random citation model.
\begin{itemize}
\item At the beginning there is one ``seminal'' paper.
\item Every sequential paper makes at most ten citations (or cites all 
preceding papers if their number does not exceed ten). 
\item All preceding papers have  an equal probability to be cited.
\item Multiple citations are excluded. So if some paper is selected 
by chance
as an citation candidate more than once, the selection is ignored (in this 
case total number of citations in a new paper will be less than ten).
\end{itemize}

\noindent
I doubt about monkeys but it is simple to learn computer to simulate such
a process. Here is the result of simulation for 1000 papers.
\newpage 
\begin{figure}[htb]
  \begin{center}
\mbox{\epsfig{figure=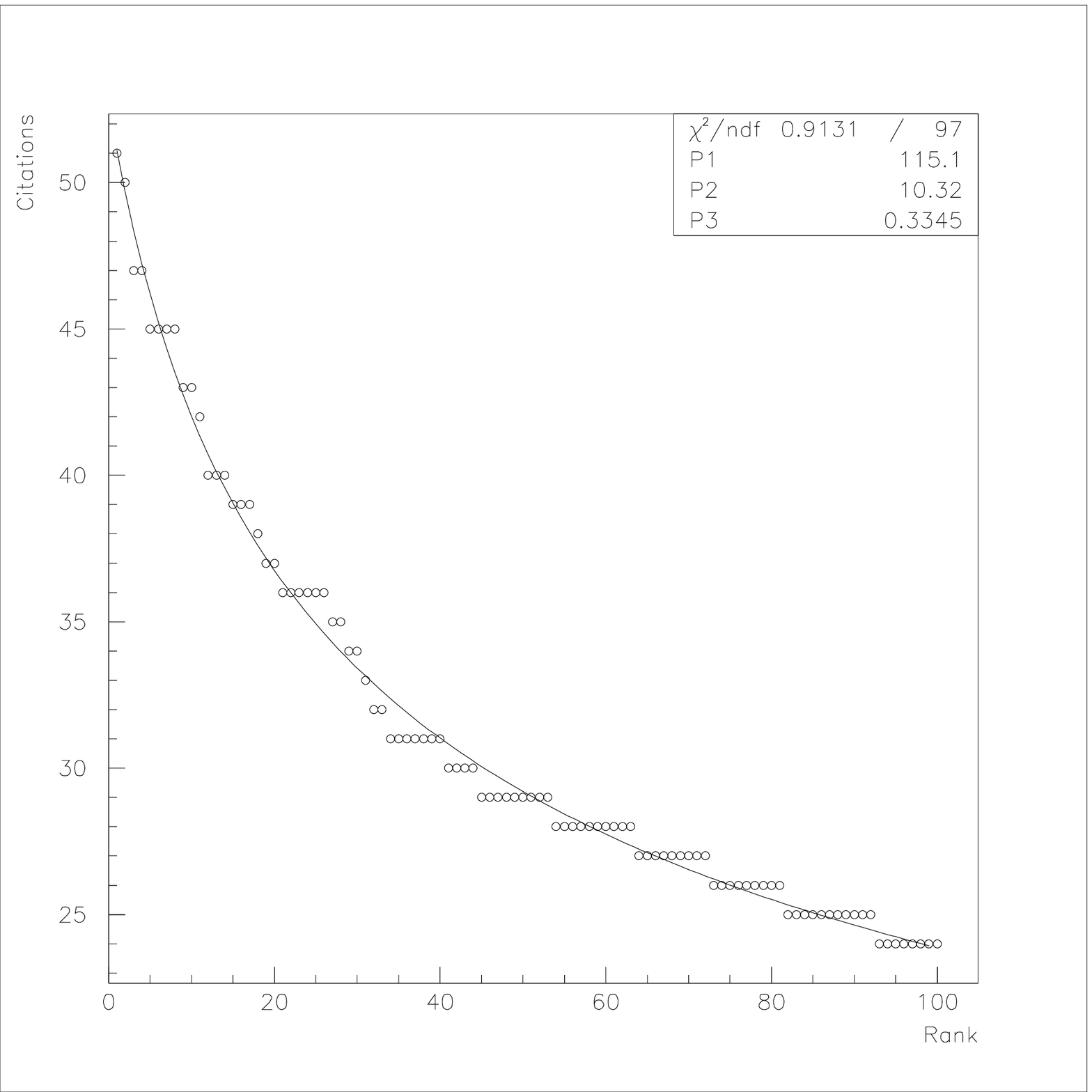
                               ,  height=14.0cm}}
%\caption {Random citations -- equiprobable}
  \end{center}
\label{Fig9}
\end{figure}

So we see an apparent pow-like structure, although with staircase behavior.
We expect this stepwise structure to disappear if we eliminate the democracy
between papers and make some papers more probable to be cited.

Note that even the value of exponent $p_3$ is reasonably close to what was
really observed for the most cited papers. But this can be merely an accident
and I do not like to make some farfetched conclusion about the nature of
citation process from this fact.

In reality ``Success seems to attract success'' \cite{9}. Therefore, 
let us try to
see what happens if the equal probability axiom is changed by perhaps a more
realistic one:
\begin{itemize}
\item The probability for a paper to be cited is proportional to $n+1$,
where $n$ is the present total citation number for the paper.
\end{itemize}

\noindent 
It is still assumed that all preceding papers compete to be cited by a new
paper, but with probabilities as follows from the above given law. The 
result for 1000 papers now looks like
\begin{figure}[htb]
  \begin{center}
\mbox{\epsfig{figure=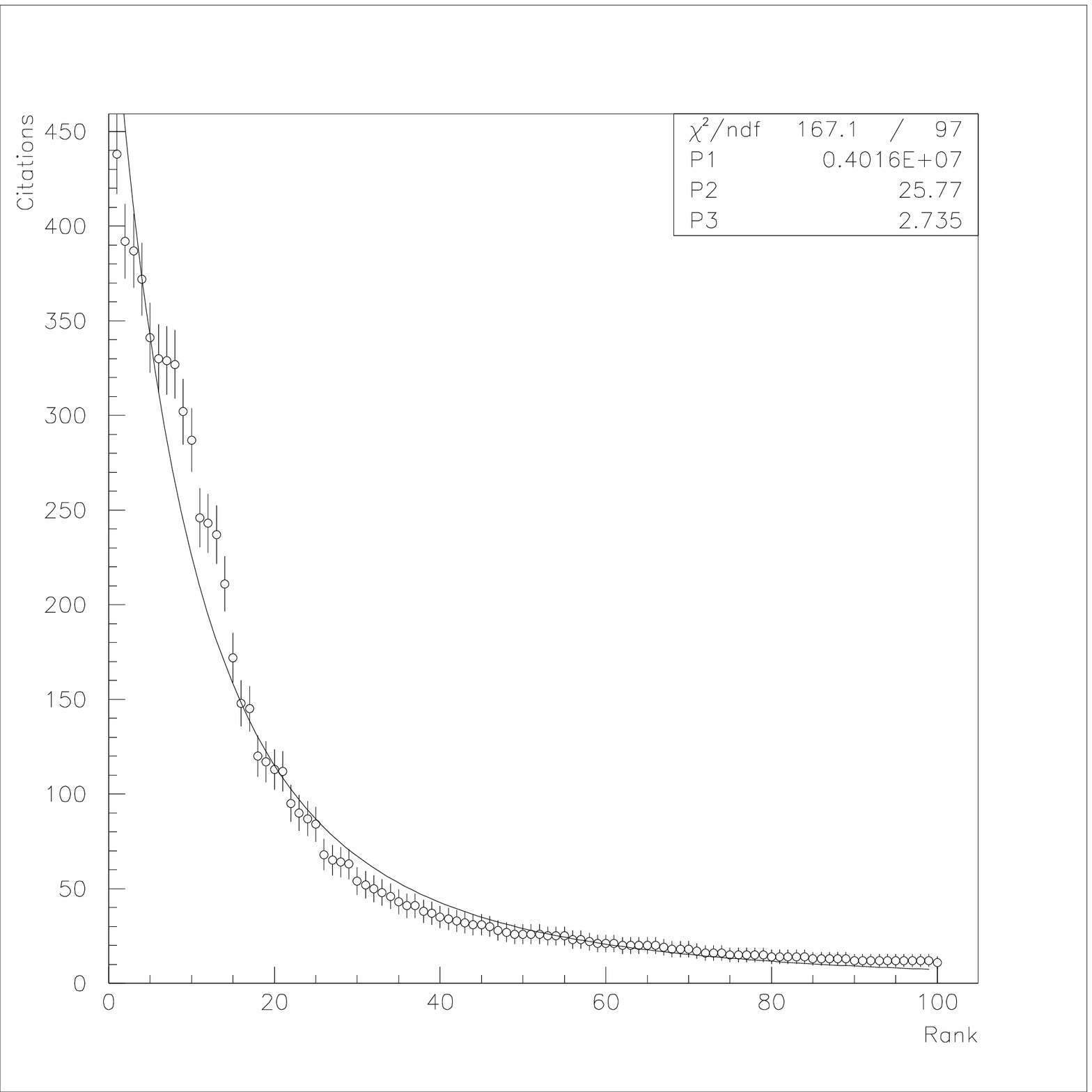
                               ,  height=14.0cm}}
   \end{center}
%\caption {Random citations -- probability ~ (citations +1)}
\label{Fig10}
\end{figure}

The fit seems not so good now, nevertheless 
you can notice some resemblance with the case of individual scientists.
Again I refrain from premature conclusions. Although it is not entirely
surprising that the  well-known a given paper of a certain author is,
the more probable becomes its citation in a new paper. 

\section{Discussion}
So scientific citations (leaving aside first name frequencies) provides
one more example of Zipf-Mandelbrot's regularity. I do not know whether
this fact indicates only to significant stochastic nature of the process
or to something else. In any case SPIRES, and the World Wide Web in general,
gives us an excellent opportunity to study the characteristics of the 
complex process of scientific citations.

I do not know either whether Mandelbrot's parameters are meaningful in this
case, and if they can tell us something non-trivial about the citation 
process.

The very generality of the Zipf-Mandelbrot's regularity can make it rather
``shallow''. But remember, that the originality of answers on the question
of whether there is something serious behind the Zipf-Mandelbrot's law depends
how restrictive frameworks we assume for the answer. Shallow framework will
probably guarantee shallow answers.  But if we do not restrict our 
imagination from the beginning, answers can turn out to be quite 
non-trivial.
For example, fractals and self-similarity are certainly great and 
not shallow
ideas. This point is very well illustrated by the ``Barometer Story'',
which I like so much that I'm tempted to reproduce it here (it is reproduced
as given in M.~Gell-Mann's book \cite{21}).  

\section{The Barometer Story -- by Dr. A.~Calandra}
Some time ago, I received a call from a colleague who asked if I would be
the referee on the grading if an examination question. It seemed that he
was about to give a student a zero for his answer to a physics question,
while the student claimed he should receive a perfect score and would do 
so if the system were not set up against the student. The instructor and
the student agreed to submit this to an impartial arbiter, and I was 
selected...

I went to my colleague's office and read the examination question, which
was, ``Show how it is possible to determine the height of a tall building
with the aid of a barometer.''

The student's answer was, ``Take the barometer to the top of the building,
attach a long rope to it, lower the barometer to the street, and then bring
it up, measuring the length of the rope. The length of the rope is the
height of the building.''

Now this is a very interesting answer, but should the student get credit 
for it? I pointed out that the student really had a strong case for full
credit, since he had answered the question completely and correctly. On
the other hand, if full credit were given, it could well contribute to a 
high grade for the student in his physics course. A high grade is supposed
to certify that the student knows some physics, but the answer to the 
question did not confirm this. With this in mind, I suggested that the 
student have another try at answering the question. I was not surprised
that my colleague agreed to this, but I was surprised that the student did.

Acting in the terms of the agreement, I gave the student six minutes to
answer the question, with the warning that the answer should show some 
knowledge of physics. At the end of five minutes, he had not written
anything. I asked if he wished to give up, since I had another class to
take care of, but he said no, he was not giving up, he had many answers
to this problem, he was just thinking of the best one. I excused myself
for interrupting him to please go on. In the next minute, he dashed off
his answer, which was: ``Take the barometer to the top of the building,
and lean over the edge of the roof. Drop the barometer, timing its fall
with a stopwatch. Then, using the formula $ s=at^2/2$, calculate the 
height of the building.''

At this point, I asked my colleague if he would give up. He conceded
and I gave the student almost full credit. In leaving my colleague's
office, I recalled that the student had said that he had other answers
to the problem, so I asked him what they were.

``Oh, yes,'' said the student. ``There are many ways of getting  the height 
of a tall building with the aid of a barometer. For example, you could take
the barometer out on a sunny day and measure the height of the barometer,
the length of its shadow, and the length of the shadow of the building,
and by the use of simple proportion, determine the height of the building.''

``Fine,'' I said. ``And the others?''

``Yes'', said the student. ``There is a very basic measurement that you will
like. In this method, you take the barometer and begin to walk up the stairs.
As you climb the stairs, you mark off the length and this will give you the
height of the building in barometer units. A very direct method.''

``Of course, if you want a more sophisticated method, you can tie the 
barometer to the end of a string, swing it as a pendulum, and determine the
value of $g$ at the street level and at the top of the building. From the
difference between the two values of $g$, the height of the building can,
in principle, be calculated.''

Finally, he concluded, ``If you don't limit me to physics solution to this
problem, there are many other answers, such as taking the barometer to the
basement and knocking on the superintendent's door. When the superintendent
answers, you speak to him as follows:

Dear Mr. Superintendent, here I have a very fine barometer. If you will tell
me the height of this building, I will give you this barometer ...''

\section*{acknowledgments}
This work was done while the author was visiting Stanford Linear Accelerator
Center. I'm  grateful to Helmut Marsiske and Lynore Tillim for kind 
hospitality.

\section*{Note added}
After this paper was completed and submitted to e-Print Archive, I have 
learned that the Zipf's distribution in scientific citations was discovered 
in fact earlier by S.~Redner \cite{41}. He also cites some previous studies
on citations, which were unknown to me.

I also became aware of G.~Parisi's interesting contribution \cite{42} from
Dr. S.~Juhos.

I thank S.~Redner and S.~Juhos for their correspondence.

\end{document}